\documentclass[12pt]{article}
\usepackage{amssymb,amsmath,graphicx}
\begin{document}
\title{\textbf{New Flavor Interactions at the LHC}} 
\author{B. Holdom%
\thanks{bob.holdom@utoronto.ca}\\
\emph{\small Department of Physics, University of Toronto}\\[-1ex]
\emph{\small Toronto ON Canada M5S1A7}}
\date{}
\maketitle
\begin{abstract}
For a new family-nonuniversal gauge interaction to be accessible at the LHC it will most likely couple preferentially to the third family. By coupling to all members of the third family the production of a new gauge boson (the $X$ with $M_X\approx1$ TeV) will lead to final states with a distinctive $\tau^+\tau^-$ pair. We study the mass reconstruction of the $X$ and the cuts that can enhance signal to background. The $X$ boson should be associated with the physics of flavor and in the simplest picture a fourth family. We discuss how the mass mixing between the third and fourth families affects the $X$ couplings and a possible mixing with the $Z$.
\end{abstract}

\section{Introduction}
The LHC is expected to shed light on the origin of electroweak symmetry breaking. There is less expectation with regard to the flavor puzzle and the origin of the quark and lepton mass spectrum. This is because the latter is usually encoded in a set of Yukawa couplings whose origins are typically expected to lie far beyond the scale of electroweak symmetry breaking. In this sense the belief in the existence of an elementary scalar field, the Higgs boson, drastically reduces our hopes for new understanding about flavor.

The picture is very different in the absence of elementary scalar fields. In this case the physics of flavor will likely have to be understood in terms of the dynamics of gauge theories.\footnote{There is also interest in the dynamics of extra dimensions, but it appears that these models require the same sort of parameterization of flavor and a fundamental lack of predictiveness as the original set of Yukawa couplings. This may be a feature of any attempt at a weak coupling description of flavor.} After all, nature provides an example of the dynamical breaking of chiral symmetries and the generation of mass with QCD. If we take the hint then we should consider the possibility of gauged flavor symmetries which are dynamically broken. In addition to the broken electroweak gauge symmetries, we have in mind more badly broken gauge symmetries that connect different families to each other. These interactions can serve to feed down mass from the heaviest fermions to the lighter ones. The physics of flavor will be characterized by scales that range from a TeV up to at least several hundred TeV, and these scales will be in some inverse relation to the masses of the fermions that are generated by the respective interaction.\footnote{The form of this inverse relationship may be affected by anomalous scaling.} Besides the small masses that the light quark and leptons receive, other effects of new flavor physics on the lightest fermions are highly suppressed. Many such effects are characterized by effective 4-fermion operators, but in the case of the three light neutrinos their masses may have to arise from 6-fermion operators. The existence of neutrino masses in the sub-eV range then turns out to be an independent indication of flavor physics below 1000 TeV.\footnote{When a Higgs is present the neutrino mass operator has dimension 5, in contrast to dimension 9 here. This explains the vast difference in the suggested scales of flavor physics at which these operators must originate.}

The lightest of the new interactions couples to the heaviest fermions and in this case the mass scale of the interaction and the fermion may be similar. In addition the heaviest fermion masses should be expected to serve as the primary order parameters for electroweak symmetry breaking.\footnote{Any heavier fermions would have to be electroweak singlets, such as the right-handed neutrinos.} The most trivial way to accommodate this union of flavor and EWSB is to extend the known flavor structure of the standard model by adding a sequential fourth family, which will contain these heaviest fermions. We suppose then that the last remaining remnant of the flavor gauge symmetry exists down to roughly a TeV. We assume that this flavor interaction is strong and that it plays some role in electroweak symmetry breaking. There may be no need for a new unbroken gauge symmetry underlying EWSB, such as technicolor. Such a picture would constitute the most economical joining of flavor and EWSB physics, and thus it is interesting to consider its experimental implications.

Clearly, the lightest of the flavor gauge bosons should couple to the fourth family. We will focus on such a gauge boson associated with a diagonal generator of some original flavor gauge group, and call it the $X$ boson \cite{a3}. Of most interest to us is the likely coupling of the $X$ boson to the third family as well. The main reason for this coupling is that this is the simplest way to cancel the gauge anomalies involving the $X$ and the electroweak gauge bosons, by having the $X$ charges of the third and fourth families being equal and opposite. This is another link between four families and the $X$ boson. With such couplings it is nevertheless still possible that the complete flavor dynamics results in only one of the two heavy families receiving the main electroweak symmetry breaking mass. The top quark mass must reflect the presence of $SU(2)_R$ breaking physics at a higher scale, and a particular 4-fermion operator can feed mass from the $b'$ to the $t$ while reducing the impact on the $T$ parameter  \cite{a1,a7}.

We have thus sketched a particular justification for the existence of a new massive gauge boson coupling to the third family.\footnote{There are also small couplings to lighter families arising from CKM related mass mixing effects. Such flavor changing couplings are probably less constrained if they occur more in the up sector.} Our assumption of the simplest possible flavor structure, along with the simplest cancellation of gauge anomalies, implies that the $X$ boson couples with equal strength $g_X$ to all members of the third family. The fact that it couples to both the $b$ quark and the $\tau$ will lead to our search strategy. Thus our analysis will apply to any other scenario having a new massive gauge boson with this basic property. We stress that the observation of such a gauge boson, nearly decoupled from the two lightest families, would open up a new window onto the physics of flavor. Although there has been much attention focussed on signals of new massive gauge bosons (often referred to as $Z'$'s) at the LHC, these studies usually involve family universal couplings. Our interest is instead with nontrivial flavor physics that is connected with origin of the sequential family structure of nature, and for this we need to see family nonuniversal physics.

To calculate the width of the $X$ boson we assume for now that it decays by pair production to all members of the third family, and that it is below the threshold for pair production of fourth family fermions. The fourth family quarks can be expected to have masses roughly of order 600 GeV.\footnote{The fourth family neutrino and charged lepton may be somewhat lighter. For constraints on the fourth family masses see \cite{a2,a7}.} Under this assumption the $X$ width is
\begin{equation}
\Gamma_X\approx g_X^2 \left[\frac{M_X}{500 \mbox{ GeV}}\right] 60 \mbox{ GeV}
.\end{equation}
The quantity we choose to fix in our study is the ratio $g_X/M_X=1/(700\mbox{ GeV})$, since this basically determines the size of possible low energy effects of the $X$ (e.g.~see Section 3). With $g_X/M_X$ fixed we shall consider the values for $M_X$ in the range from 700 to 1300 GeV. The corresponding couplings are quite strong (these are couplings renormalized at the scale $M_X$) and the $X$ width can become quite large, ranging from 84 to 540 GeV.

\section{Signal and Background}
$X$ is produced in $pp$ collisions due to its coupling to $b$ quarks. Due to the bottom sea quark component of the protons, the main parton-level processes in order of importance at the LHC are the following.
\begin{eqnarray*}
b\overline{b}&\rightarrow&X\\
g(b\mbox{ or }\overline{b})&\rightarrow&X g(b\mbox{ or }\overline{b})\\
gg&\rightarrow&X b\overline{b}\\
q(b\mbox{ or }\overline{b})&\rightarrow&X q(b\mbox{ or }\overline{b})\quad(q=\mbox{light quark})
\end{eqnarray*}
The first process contributes about 2/3 of the cross section and the second about 1/4. We focus on the decay mode $X\rightarrow\tau^+\tau^-$.

The reconstrution of the $X$ mass can be accomplished \cite{a4} even though the $\tau$ decays produce missing energy. The point is that the $\tau$'s are highly boosted and so the missing component of their decay products are close to collinear with the visible components.\footnote{The accuracy of this approximation is more than adequate, due to the large natural width of the $X$.} We can write the $\tau^\pm$ momenta as $\vec{p}_+/x_+$ and $\vec{p}_-/x_-$, where $x_+$ and $x_-$ are the fractions carried by the visible components $\vec{p}_+$ and $\vec{p}_-$. Then the measured total transverse missing momentum is
\begin{equation}
\vec{p}\!\!\!/_T=(\frac{1}{x_+}-1)\vec{p}_{T+}+(\frac{1}{x_-}-1)\vec{p}_{T-}
.\label{e1}\end{equation}
This vector relation determines $x_+$ and $x_-$. Then the $X$ invariant mass is obtained by scaling up the invariant mass determined by the four-vectors $p_+$ and $p_-$ by a factor of $1/\sqrt{x_+x_-}$.

For background event generation we use the Alpgen\cite{B}-Pythia\cite{A} combination, while for signal generation we use MadEvent\cite{C}-Pythia to accommodate the new physics model. We adopt the Pythia tune D6T (due to R.~Field) which is based on the CTEQ6L1 PDF.\footnote{Pythia is run with $\tau$ decays treated internally and with $\pi^0$'s treated as stable.} Since we do not go beyond tree-level matrix elements in our event generation, it is important to choose a sensible renormalization/factorization scale that can be used consistently for both signal and background. We choose to set this scale to the partonic quantity $\sqrt{\hat{s}}/2$ (which goes over to the canonical choice of $m_t$ for $t\overline{t}$ production close to threshold). We note that other popular choices for the renormalization scale involve some combination of $p_T$'s and masses of the particles produced, but in our case the mass of the $X$ is not well defined when it is a very broad resonance.

MadEvent is run for the combined set of processes listed above with a constraint that the jets, including $b$-jets, are required to have $p_{T}>20$ GeV. This is due to the issue of double-counting; if the bottom sea quarks in the proton are thought to arise from gluon splitting $g\rightarrow b\overline{b}$, then the first process above could be thought to have the same topology as a diagram contributing to the third process. The overlap is expected to occur for $b$-jets at small $p_{T}$ and large $\eta$, and hence the $p_{T}>20$ GeV cut. With this cut the third process in total contributes less than 5\% of the cross section and thus any remaining double-counting cannot be significant. We find cross sections for $X$ production (in collisions of 7 TeV protons) of 4.4, 2.8, 1.9, 1.4, 1.1 pb for $M_X=700, 850, 1000, 1150, 1300$ GeV (with $g_X=M_X/(700\mbox{ GeV})$).

Alpgen generates background events with MLM jet matching involving 0, 1, and 2 extra hard jet samples, with jets having $p_{T}>50$ GeV and pseudorapidity $|\eta|<2$. PGS4\cite{D} is used for detector simulation, with the parameter choice for ATLAS as supplied by the MadEvent package.

For event selection we use the following cuts, where by `lepton' we mean an isolated electron or muon or a $\tau$-tagged jet. The $\tau$-tagged jets are in fact essential for the signal.\footnote{A restriction to 1-prong $\tau$-tags is also worth considering, which could give comparable results.}
\begin{itemize}
\item at least one pair of oppositely charged leptons, each with $p_T>60$ GeV, with invariant mass $>300$ GeV
\item missing energy $p_T\!\!\!\!\!\!/\;\;>60$ GeV
\item $H_T>700$ GeV
\item not more than one non-$b$-tag jet with $p_T>60$ GeV
\item no jets or leptons with $|\eta|>2$ and $p_T>60$ GeV
\end{itemize}

Very effective is the 300 GeV dilepton invariant mass constraint, which essentially eliminates the contribution from the $(Z\rightarrow\ell^+\ell^-)+$jets background. To reconstruct the $X$ mass we consider all such pairs of leptons and require that the $x_+$ and $x_-$ obtained by solving (\ref{e1}) satisfy $0<x_\pm<1$. Also effective is the requirement of not more than one non-$b$-tag jet to suppress backgrounds involving more jets, which includes $t\overline{t}+$jets. A low rate for jets to fake $\tau$'s, along with the missing energy constraint, will help to suppress the pure QCD jet background. The missing energy constraint could be increased further if necessary. We will take the $\tau$ fake rate conservatively to be 1\% (lower values are described in \cite{a5} for ATLAS).

We are left with $t\overline{t}+$jets and $W+$jets as the main backgrounds. In our generation of events for the former we take both top quarks to decay semileptonically, and for the $W+$jets background we take the $W$ to decay leptonically. For the $W+$jets background, at least one of the two leptons used to reconstruct the $X$ will be a jet that fakes a lepton (most likely a fake $\tau$). The PGS4 $\tau$ fake rate appears to be too large, and to implement our own fake rate we proceed as follows. We consider another set of events which has an additional jet which is neither $b$-tagged nor $\tau$-tagged which we use along with an identified lepton in the event to reconstruct the $X$ mass. By combining this set of events with the original set we then have both PGS4 $\tau$-tagged (fake) and non-tagged jets contributing to the reconstruction,\footnote{The contribution from the non-tagged set is divided by 2 since the charge of the fake will not be correct half of the time.} and to this result we can multiply by our estimate of a $\tau$ fake rate (1\%). For the signal and the $t\overline{t}+$jets background there are two real leptons in the events and in these cases we don't make any correction for the PGS4 $\tau$ fake rate.

It is also useful to consider the relative orientation of the three vectors $\vec{p}_{T+}$, $\vec{p}_{T-}$ and $\vec{p}\!\!\!/_T$ appearing in (\ref{e1}). In particular consider the vector that bisects the angle between $\vec{p}_{T+}$ and $\vec{p}_{T-}$. Let $\phi_1$ be the angle between this vector and $\vec{p}\!\!\!/_T$. Then the signal favors  $|\sin(\phi_1)|\approx 1$ more so than the background. This is due to the fact that $\vec{p}_{T+}$ and $\vec{p}_{T-}$ tend to be close to back-to-back, in both signal and background, while for the signal the missing momentum from each $\tau$ decay aligns with $\vec{p}_{T+}$ and $\vec{p}_{T-}$ respectively. Thus a cut $|\sin(\phi_1)|>2/3$ for example will suppress the background more than the signal, especially for larger $M_X$. We shall impose this cut, while noting that other related cuts can also be devised.

\begin{center}\includegraphics[scale=0.65]{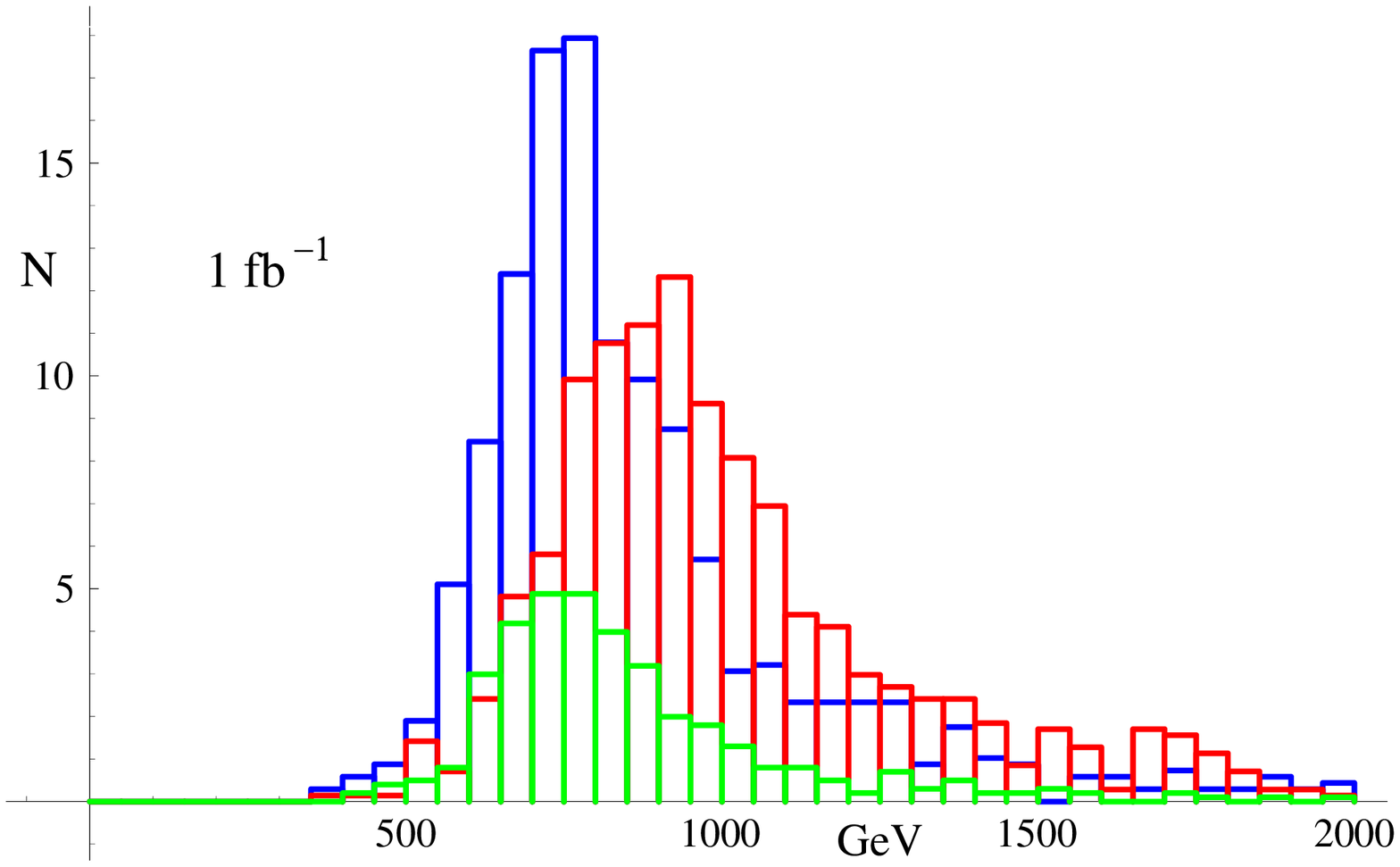}
\end{center}
\vspace{-1ex}\noindent Figure 1: The signal for $(g_X, M_X[GeV])$:  (1, 700, blue), (1.21, 850, red), (0.5, 700, green). The numbers of events are normalized to 1 fb$^{-1}$.
\vspace{2ex}
\begin{center}\includegraphics[scale=0.65]{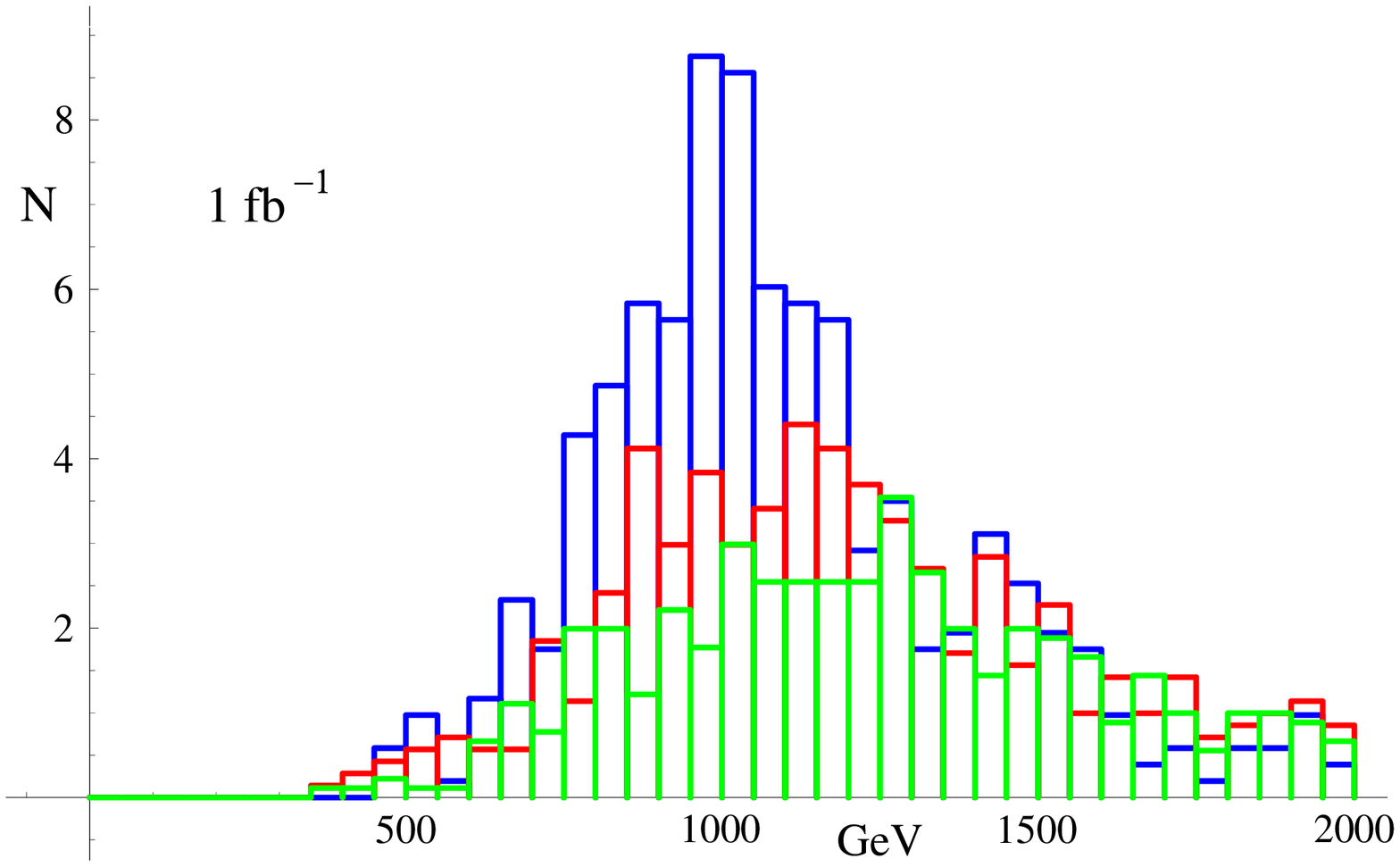}
\end{center}
\vspace{-1ex}\noindent Figure 2: The signal for $(g_X, M_X[GeV])$:  (1.43, 1000, blue), (1.64, 1150, red), (1.86, 1300, green). The vertical scale differs from the previous figure.
\vspace{2ex}

The $X$ has a production rate that falls rapidly with its mass, and a width that is proportional to $g_X^2M_X$. Both of these effects cause its detectability at the LHC to fall with increasing mass even while $g_X/M_X$ is held fixed.  We show the $X$ mass reconstruction in Figs.~(1,2) for $M_X=700, 850, 1000, 1150, 1300$ GeV for $g_X=M_X/(700\mbox{ GeV})$. In Fig.~(1) we also display the case of a smaller coupling, $g_X=0.5$ for $M_X=700$ GeV. By comparing these results one sees how both the height and width of the peak provides information on $g_X$. Fig.~(3) shows that our cuts have been effective at reducing the backgrounds to levels small compared to the signal. Even a high mass broad signal peak can be differentiated from the background since the background peaks around 700 GeV.
\begin{center}\includegraphics[scale=0.65]{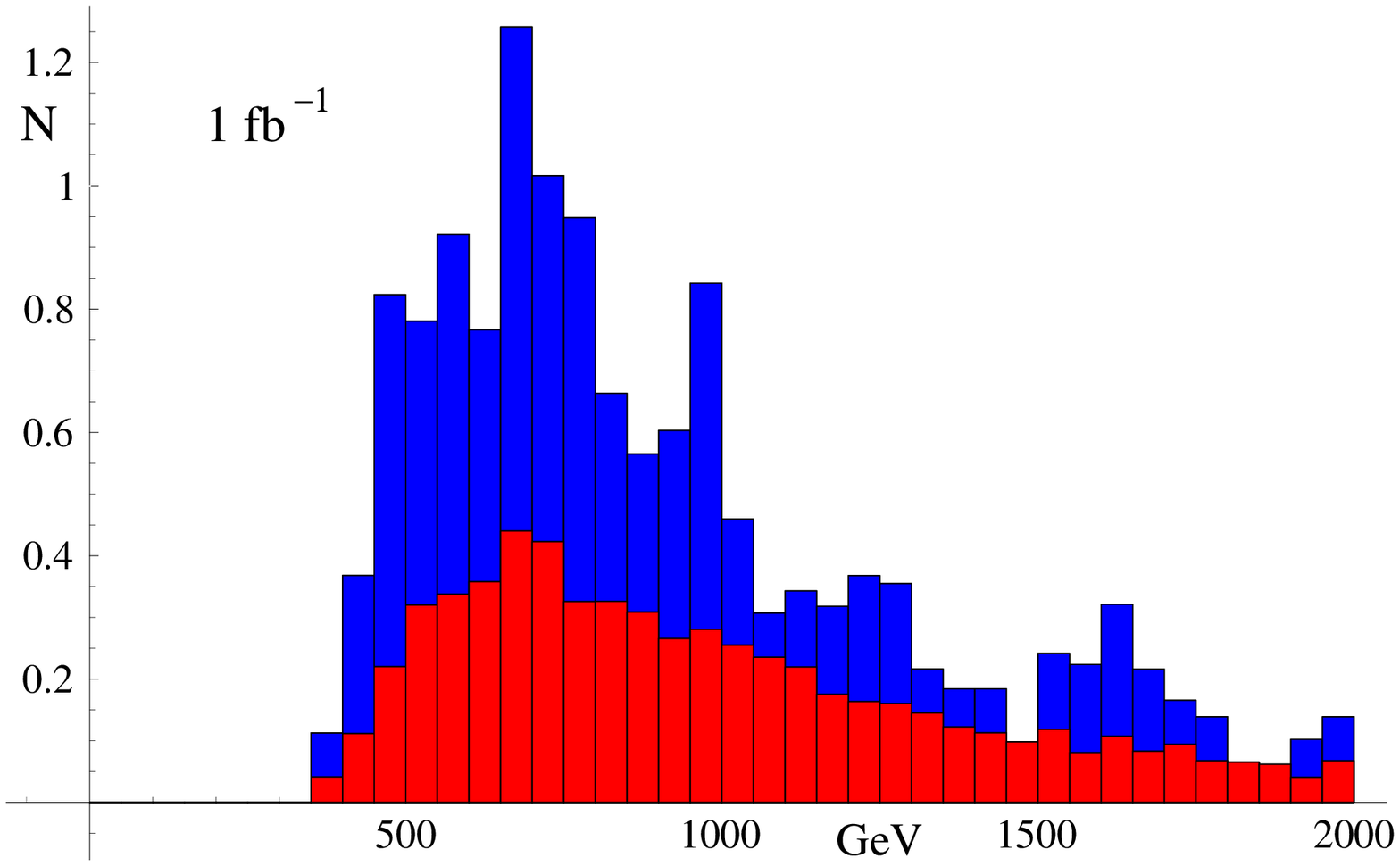}\end{center}
\vspace{-1ex}\noindent Figure 3: The background for $t\overline{t}+$jets (blue) and $W+$jets (red). These histograms are stacked, not overlayed. The vertical scale differs from the previous figures.
\vspace{2ex}

The $X$ boson should be easy to differentiate from a $Z'$ boson with family universal couplings. Through its coupling to light quarks a $Z'$ can be found up to significantly higher mass. And a $Z'$ with a mass of order a TeV is constrained to have couplings typical of the $Z$ or smaller and would thus be much more narrow than we expect for the $X$ boson. To test the hypothesis that the $X$ couples equally to all members of the third family, we need to consider other decay modes of the $X$ boson.

Decays to a pair of quarks can produce $b\overline{b}b\overline{b}$, $b\overline{b}t\overline{t}$ and $t\overline{t}t\overline{t}$ final states, but the extraction of these signals from background appears to be nontrivial and requires more study. The decay to $\nu_\tau\overline{\nu}_\tau$ can lead to $b\overline{b}+/\!\!\!\!E_T$, where the missing transverse momentum is that of the $X$. But the latter is not expected to be large compared to the $X$ mass, and the result is that the signal is swamped by the $Z+$jets background where $Z\rightarrow\nu\overline{\nu}$ and the jets fake $b$-jets. More promising is the decay to the fourth neutrino $X\rightarrow\nu_{\tau'}\overline{\nu}_{\tau'}$ if kinematically allowed. Since mass mixing in the lepton sector can lead to the decay $\nu_{\tau'}\rightarrow (\tau,\mu\mbox{ or }e) +W$, this can lead to interesting signatures involving varying numbers of leptons and jets. Of course $\nu_{\tau'}\overline{\nu}_{\tau'}$ could also be produced directly through a $Z$.

\section{A fourth family and mixing effects}
We first discuss how mass mixing between the third and fourth families can affect the $X$ couplings to quarks. Let us consider for example the $b$ and $b'$. In an interaction eigenstate basis with $\hat{B}=(\hat{b}',\hat{b})$ the $X$ couples with coupling strength $g_X$ to a vector current
\begin{equation}
J_\mu=\overline{\hat{B}}_L\gamma_\mu \hat{Q} \hat{B}_L+\overline{\hat{B}}_R\gamma_\mu \hat{Q} \hat{B}_R\quad \hat{Q}=\left(\begin{array}{cc}1 & 0 \\0 & -1\end{array}\right)
.\end{equation}
But the $(\hat{b}',\hat{b})$ fields need not correspond to the mass eigenstates. The unitary transformations $B_L=U_L \hat{B}_L$ and $B_R=U_R \hat{B}_R$ to the mass eigenstate basis produce a nondiagonal current
\begin{equation}
J_\mu=\overline{B}_L\gamma_\mu Q_L B_L+\overline{B}_R\gamma_\mu Q_R B_R\quad Q_L=U_L\hat{Q}U_L^\dagger\quad Q_R=U_R\hat{Q}U_R^\dagger
\end{equation}
The third family $Xb_L\overline{b}_L$ and $Xb_R\overline{b}_R$ couplings are then determined by the respective components of the matrices $Q_L$ and $Q_R$. These diagonal components vary between $-1$ and 1, and thus the $Xb_L\overline{b}_L$ and $Xb_R\overline{b}_R$ couplings independently vary between $-g_X$ and $g_X$. The off-diagonal left-handed couplings $Xb_L\overline{b}_L'$ and $Xb_L'\overline{b}_L$ have absolute value $\le g_X$ and they only vanish when the $Xb_L\overline{b}_L$ couplings are $\pm g_X$. The same is true for the right-handed couplings.

The same story applies to the ($t'$,$t$), ($\tau'$,$\tau$) and ($\nu'_\tau$,$\nu_\tau$) pairs, with potentially different unitary transformations. But we should assume that the $X$ couplings preserve to good approximation the usual custodial $SU(2)$ symmetry among $t'$ and $b'$ quarks. This implies that we can write the couplings to the third family mass eigenstates in terms of four parameters (fourth family couplings are opposite in sign).
\begin{eqnarray}
g_L^{Xt}=g_L^{Xb}\equiv g_L^{Xq}&&\\
g_R^{Xt}=g_R^{Xb}\equiv g_R^{Xq}&&\\
g_{L}^{X\nu}=g_{L}^{X\tau}\equiv g_{L}^{X\ell}&&\\
g_{R}^{X\tau}\equiv g_{R}^{X\ell}
\end{eqnarray}
If the $X$ boson has some axial quark coupling $g^{Xq}_A=(g_L^{Xq}-g_R^{Xq})/2\neq0$ then it will receive a contribution to its mass from the dynamical $t'$ and $b'$ masses. Given that essentially all of the $Z$ mass arises from this source we have the bound
\begin{equation}
\frac{(g^{Xq}_A)^2}{M_X^2}\le\frac{\left(\frac{e}{4cs}\right)^2}{M_Z^2}
.\label{e3}\end{equation}

An axial coupling can also imply some mass mixing between the $X$ and the $Z$ \cite{a3}. Let us denote this by $M_{ZX}^2$, the off-diagonal element of the $2\times2$ mass-squared matrix. The $t'$ and $b'$ contributions cancel in the limit of $t'$-$b'$ mass degeneracy, but a $t$ loop contributes as long as $g^{Xq}_A\neq0$. We can write this contribution to $M_{ZX}^2$ in terms of $f_t$, which is defined such that $f_t^2/v^2$ with $v\approx240$ GeV gives the fractional contribution of the $t$-loop to $M_Z^2$,
\begin{equation}
M_{ZX}^2=\frac{e}{cs}g^{Xq}_A f_t^2
.\end{equation}
A standard estimate is $f_t\approx60$ GeV \cite{a6}. Fourth family leptons could also contribute to this mixing, but as we shall see the $\tau'$ must have mostly vectorial couplings to the $X$ and the $\nu_{\tau'}$ mass is probably not large compared to the $t$. In terms of the mass mixing $M_{ZX}^2$ we have a shift of the $Z$ couplings to the third family
\begin{equation}
\delta g^f_{L,R}=-\frac{M_{ZX}^2}{M_X^2}g_{L,R}^{Xf}\quad f=q,\ell
.\label{e2}\end{equation}

The current constraints from $\delta\Gamma_b/\Gamma_b$ and $\delta\Gamma_\tau/\Gamma_\tau$ are quite strong and imply that substantial shifts in the $Z$ couplings are only allowed if the shift in the $b$ coupling is mostly right-handed ($\delta\Gamma_b=0$ if $\delta g^q_{L}=0.18\delta g^q_{R}$) and the shift in the $\tau$ coupling is mostly vectorial ($\delta\Gamma_\tau=0$ if $\delta g^\ell_{L}=0.86\delta g^\ell_{R}$). Since the shifts (\ref{e2}) are proportional to the $X$ couplings, for illustration we consider the following choices
\begin{equation}
g_L^{Xq}=0,\quad g_R^{Xq}=-g_L^{X\ell}=-g_R^{X\ell}=\hat{g}_X
.\end{equation}
This will essentially give the maximal shifts in quantities other than $\delta\Gamma_b/\Gamma_b$ and $\delta\Gamma_\tau/\Gamma_\tau$.

Also for illustration we will consider the case where $\hat{g}_X/M_X$ is chosen to produce results similar to our Monte Carlo simulations where $g_X/M_X=1/(700\mbox{ GeV})$. For this we need to set $\hat{g}_X=2^\frac{1}{4}g_X$  to make up for the absence of the coupling to $b_L$.\footnote{We ignore a possible increase of the width of the $X$ if the decay to $b'\overline{b}$ etc.~became possible.} The resulting $(g_A^{Xq})^2/M_X^2$ is safely compatible with the bound in (\ref{e3}). Combining these values with the above results we obtain the following shifts in the $Z$ couplings to the third family,
\begin{equation}
\delta g^q_L=0,\quad \delta g^q_R=-\delta g^\ell_L=-\delta g^\ell_R\approx0.00386
.\end{equation}

We compare the resulting pattern of deviations to the experimental results \cite{a9} in Table 1. The experimental $\delta\Gamma_b$ and $\delta A_b$ values were obtained from the departures from the standard model fit prediction. $\delta\Gamma_\tau$ and $\delta A_\tau$ were derived from the measured departures from lepton universality, assuming $\Gamma_e=\Gamma_\mu$ and $A_e=A_\mu$. $\delta\Gamma_{\nu_\tau}$ was attributed to the departure of the measured number of neutrinos from three. We see that the $X$ boson induced shifts are compatible with experiment except for two of the measurements of the $\tau$ asymmetry parameter. This apparently rules out this picture of maximal $Z$ coupling shifts unless some experimental errors are seriously underestimated. The $X$ boson induced shifts can be continuously decreased to zero by increasing the size of the vector relative to the right-handed $X$ couplings to quarks.

\begin{center}\begin{tabular}{|l|l|l|}\hline  & Experimental & $X$ boson \\\hline  $\delta\Gamma_b/\Gamma_b$& $0.0022\pm0.003$ & 0.004 \\\hline $\delta\Gamma_\tau/\Gamma_\tau$ & $0.0036\pm0.0025$ & 0.003  \\\hline $\delta\Gamma_{\nu_\tau}/\Gamma_{\nu_\tau}$ & $-0.016\pm0.008$ & $-0.02$\\\hline$\delta A_b/A_b$(FB) & $-0.046\pm0.016$ & $-0.009$ \\ $\delta A_b/A_b$(LR) & $-0.013\pm0.022$ & \\\hline$\delta A_\tau/A_\tau$(FB) & $0.24\pm0.17$ & \\$\delta A_\tau/A_\tau$(LR) & $-0.1\pm0.1$ &0.27 \\$\delta A_\tau/A_\tau$(${\cal P}_\tau$) & $-0.04\pm0.04$ &  \\\hline \end{tabular}\end{center}
\noindent Table 1: FB refers to the forward-backward asymmetries, LR the left-right asymmetries from SLD, and ${\cal P}_\tau$ the tau polarization measurements.

\section{Conclusions}
We have considered a picture for minimal new flavor interactions, where the $X$, a new massive gauge boson, couples preferentially to the quarks and leptons of the third family. The coupling can be of order one or more and this can lead to a large width for the $X$. The $X$ is produced at the LHC through its coupling to the $b$ quark, while the decay $X\rightarrow\tau^+\tau^-$ produces the most striking signatures. These events will be deficient in non-$b$-jets. The $X$ mass can be reconstructed from the visible decay products of the $\tau$'s and missing $p_T$. We have discussed event selection cuts that quite effectively suppress main backgrounds from $t\overline{t}+$jets and $W+$jets. This leads to a clear discovery potential for the $X$ with mass ranging up to at least a TeV, with only a fb$^{-1}$ of integrated luminosity.

We have restricted ourselves to $X$ masses at or above 700 GeV because this is roughly the scale where we could expect that the physics of flavor comes together with the physics of electroweak symmetry breaking. This picture points towards a sequential fourth family with similar mass. A fourth family also allows the required cancellation of gauge anomalies associated with a $X$ boson. A heavy fourth family has its own signatures at the LHC \cite{a7}, and in particular $t'\overline{t}'$ production provides a signal for which $t\overline{t}$ production is a dominant irreducible background, just as we have found for $X$ production. This indicates that a thorough understanding of the $t\overline{t}$ sample at the LHC will be a prerequisite in the search for new flavor physics.

\section*{Acknowledgments}
I thank W.~Bernreuther for a useful discussion. This work was supported in part by the Natural Science and Engineering Research Council of Canada.

\end{document}